\documentclass[11pt,titlepage]{article}
\usepackage{epsf}
\usepackage[dvips]{graphicx}
\usepackage{psfrag}
\newcommand{\be}{\begin{equation}}
\newcommand{\ee}{\end{equation}}
\newcommand{\bea}{\begin{eqnarray}}
\newcommand{\eea}{\end{eqnarray}}
\def\lsim{\raise0.3ex\hbox{$\;<$\kern-0.75em\raise-1.1ex\hbox{$\sim\;$}}}
\def\gsim{\raise0.3ex\hbox{$\;>$\kern-0.75em\raise-1.1ex\hbox{$\sim\;$}}}
\newcommand{\nn}{\nonumber}

\newcommand{\AmS}{{\protect\the\textfont2
  A\kern-.1667em\lower.5ex\hbox{M}\kern-.125emS}} 
\begin{document}

\begin{titlepage}
\begin{center}
{\Large \bf $B \to \pi \pi$ decays*}

\vspace{2cm}

{\large \bf  T. N. Pham}\\

{\it Centre de Physique Th\'eorique,\\ Centre National de la Recherche 
Scientifique, UMR 7644,\\ Ecole Polytechnique, 91128 Palaiseau Cedex,
France}
\vspace{5cm}

\end{center}
\centerline{\large \bf ABSTRACT}
\vspace{0.5cm}
 The branching ratios and  CP asymmetries of 
 $B\to \pi\pi$ decays 
 measured in B factory experiments
 indicate a large ratio of color-suppressed ($C$) to
color-allowed ($T$) tree contributions and a large relative phase between
the tree and penguin amplitude. In this talk, I would like to
report on a recent analysis  to see  whether the large  $C/T$ ratio can be 
explained within the QCD based model  with i) a large contribution from the 
annihilation terms  or with ii) large 
final-state-interaction phase between two different isospin amplitudes. 
We show that the current experimental data do not exclude either
 possibility but we may be able to distinguish these two effects 
in  future measurements of direct CP asymmetry in $B\to \pi^{0}\pi^{0}$ 
decay.

\vfill

\hrule width6cm height0.5pt\hfil\break

\vskip-0.4cm

{\large *} Talk given at 
First Workshop On Theory, Phenomenology And Experiments In Heavy Flavor, 
Anacapri, Italy, May 29-31, 2006

\end{titlepage}
\section{Introduction}

The large $B\to \pi^{0}\pi^{0}$ branching ratio measured at Babar and Belle
indicates that the color-suppressed amplitude is large compared with
predictions from the short-distance effective non-leptonic weak Hamiltonian.
the ratio of the color-suppressed to color-allowed tree amplitude  $C/T$ is
 proportional to  $a_{2}/a_{1}$
with $a_{1}=C_{1} +C_{2}/N_{c}$ and $a_{2}=C_{2} +C_{1}/N_{c}$ being
the effective Wilson coefficients for the charged and neutral
current tree-level color-singlet operator $O_{1}$ and $O_{2}$
respectively. At the 
renormalization scale $\mu \approx m_{b}$\cite{Buras} , 
 $C_{1}\approx 1.10, C_{2}\approx -0.250$ (at NLO) for non-leptonic $B$
decay,    $a_{2}$ therefore becomes much smaller than  $a_{1}$. Thus 
in the absence
of higher order QCD radiative corrections and non-factorisable 
contributions, $B$ decays with neutral particles in the final state 
are suppressed compared to decays with charged particles in the final
state and are called color-suppressed decays. This rule is known to be
violated in $D$ decays and also in $B$ decays with charmed
meson in the final state. Since we expect QCD Factorization (QCDF)\cite{Beneke}
to work better for  charmless two-body $B$ decays with energetic light 
hadrons in the final state, the large $C/T$ ratio extracted from the
BaBar and Belle $B \to \pi\pi$ measurements compared with theoretical
predictions from  factorization models
is a surprise: The 
suppression of the $B^{0}\to \pi^{+}\pi^{-}$ and the enhancement of 
the $B^{0}\to \pi^{0}\pi^{0}$ decays. Also the large measured CP asymmetry 
in $B^{0}\to \pi^{+}\pi^{-}$  implies a large  relative
phase between the tree and penguin amplitude and suggests that strong
final state  interaction effect could play an important role
in $B \to \pi\pi$ decays which are governed not only by weak interaction
but also by strong interactions.
Therefore an understanding of the strong interaction effects in these
processes is crucial for extracting the weak phase and ultimately
possible new physics which could make additional 
contributions to the one-loop penguin amplitude \cite{Kou0}.
In this talk, I would like to report on a recent work\cite{Kou}
in which  we investigate two possible mechanisms for the
enhancement of $C/T$ in $B \to \pi\pi$ decays:
i) the higher order correction of QCDF and ii) the effect of final state
interaction (FSI) and show that in QCDF annihilation terms
could play an important role in the enhancement of $C/T$ but
alternatively, a large FSI rescattering phase could also 
enhance $C/T$ in QCDF with moderate annihilation contributions. 

\section{The decay amplitude from  experiments}
With the recent measurements of the  branching ratios
for the $B^{0}\to \pi^{+}\pi^{-}$, $B^{0}\to \pi^{0}\pi^{0}$ and
$B^{+}\to \pi^{+}\pi^{0}$ as well as the mixing-induced  
$S_{\pi^{+}\pi^{-}}$ and direct $C_{\pi^{+}\pi^{-}}$ CP asymmetries 
in $B^{0}\to \pi^{+}\pi^{-}$ decays, it is now pos  sible to determine the
ratio of penguin to tree amplitude and the relative phase between 
tree and penguin amplitude as done in  many previous model-independent
analysis \cite{Gronau}-\cite{Raz}.

 In the c-convention\cite{Gronau}, the $B\to \pi\pi$ amplitudes
are  
\bea
&&{\rm A} (B^0\to \pi^+\pi^-)=T e^{i\delta_T}e^{i\gamma}+P e^{i\delta_P}  \label{eq:1}\\
&&\sqrt{2}{\rm A} (B^0\to\pi^0\pi^0)=C e^{i\delta_C}e^{i\gamma}-P e^{i\delta_P} \label{eq:2}\\
&&\sqrt{2}{\rm A} (B^+\to \pi^+\pi^0)=(T e^{i\delta_T}+C e^{i\delta_C})e^{i\gamma}  \label{eq:3}
\eea
where $\delta_{\rm T,C,P}$ are respectively the strong phase for
the tree $T$, color-suppressed $C$ and penguin $P$ amplitude, $\gamma$
being the CP violating weak phase. To determine $C/T$ and the relative 
phase between $C$ and $T$, we have 5 measured 
observables \cite{HFAG} (all branching ratios quoted are CP-averaged)
 \bea
S_{\pi^+\pi^-}&=&-0.50\pm 0.12 \label{eq:data1}\\
C_{\pi^+\pi^-}&=&-0.37\pm 0.10 \\
{\rm Br}(\pi^+\pi^-)&=&(4.5\pm 0.4)\times 10^{-6} \label{eq:data3}\\
{\rm Br}(\pi^0\pi^0)&=&(1.45\pm 0.29)\times 10^{-6} \\
{\rm Br}(\pi^+\pi^0)&=&(5.5\pm 0.6)\times 10^{-6} \label{eq:data5}
 \eea 
where $S_{\pi^+\pi^-}$ and $C_{\pi^+\pi^-}$ are respectively the 
$B^{0}-\bar{B^{0}}$ mixing-induced and direct CP asymmetries in the 
expression for the time-dependent CP asymmetry of $B\to \pi^+\pi^-$:
\be
  A_{\pi^+\pi^-}(t)= S_{\pi^+\pi^-}\sin (\Delta M_B t)
-C_{\pi^+\pi^-}\cos (\Delta M_B t) 
\ee
In terms of $T$, $C$ and $P$, we have:
\be
 R\ S_{\pi^+\pi^-}\label{eq:S}
 =\sin 2\alpha
 +2\sin(\beta-\alpha)\cos\delta_{PT}\left(\frac{P}{T}\right)
 -\sin 2\beta \left(\frac{P}{T}\right)^2  
\ee
\be
R\ C_{\pi^+\pi^-}
 \label{eq:C} 
=2\sin (\alpha+\beta)\sin\delta_{PT} \left(\frac{P}{T}\right)  
  \ee 
where
 \be
 R=1-2\cos(\alpha+\beta)\cos\delta_{PT}  \left(\frac{P}{T}\right) +
  \left(\frac{P}{T}\right) ^2  \label{eq:R}
 \ee
The CP-averaged branching ratios are given by:
 \bea
&&R_{00}=\frac{2{\rm Br}(\pi^0\pi^0)}{{\rm Br}(\pi^+\pi^-)}\label{eq:rn}
=\frac{1}{R}\left[ \left(\frac{C}{T}\right)^2 + \left(\frac{P}{T}\right)^2 \right. \left. -2\cos(\delta_{PT}-\delta_{CT})\cos\gamma \left(\frac{C}{T}\right)  \left(\frac{P}{T}\right)\right]\\
&&R_{+-}=\frac{2{\rm Br}(\pi^+\pi^0)\tau_{B^0}}{{\rm Br}(\pi^+\pi^-)\tau_{B^+}} \label{eq:rc} 
 =\frac{1}{R}\left[ 1 +2\cos\delta_{CT}\left(\frac{C}{T}\right)+\left(\frac{C}{T}\right)^2  \right]
 \eea
To have an idea of the size of the relative phase $\delta_{PT}$, let
us take as an approximation,  $\alpha\approx 90^{\circ}$ and neglect
the $(P/T)^{2}$ term in Eq.(\ref{eq:S}) , we have
 \be
 \tan\delta_{PT} \approx - C_{\pi^+\pi^-}/ S_{\pi^+\pi^-}
 \ee
which gives $\delta_{PT} = -36.5^{\circ} $
more or less in agreement with the 
 more precise value $-41.3^{\circ}$ for $\gamma = 67^{\circ}$ obtained
in Table 1. This simple calculation  shows that
there is a large relative penguin-tree strong phase in 
$B\to \pi\pi$ decays. Table 1 also shows the variation of $P/T$, $R$
and $\delta_{PT}$ with $\gamma$ and  CP asymmetries within 
the uncertainties  in the measured
CP asymmetries and the  determination of $\gamma$.
\begin{table*}[t!]
\begin{center}
\begin{tabular}{|c||c|c|c|c|c|c|c|}\hline
 $\gamma$ &&\\
$(S_{\pi^+\pi^-}, C_{\pi^+\pi^-})$ &$27^{\circ}$&$37^{\circ}$&$47^{\circ}$&$57^{\circ}$&$67^{\circ}$&$77^{\circ}$&$87^{\circ}$\\
\hline\hline
 & 0.53&0.38&0.36&0.46&0.63&0.81&0.98\\
(-0.62, -0.47)&$-155^{\circ}$&$-131^{\circ}$&$-92.6^{\circ}$&$-61.7^{\circ}$&$-45.0^{\circ}$&$-35.5^{\circ}$&$-29.5^{\circ}$\\ 
& 0.43&0.75&1.10&1.45&1.75&1.96&2.06 \\ \hline 
&0.49&0.29&0.19&0.32&0.52&0.72&0.91\\
(-0.62, -0.27)&$-166^{\circ}$&$-147^{\circ}$&$-92.1^{\circ}$&$-43.4^{\circ}$&$-27.5^{\circ}$&$-20.5^{\circ}$&$-16.6^{\circ}$\\ 
&0.40&0.69&1.03&1.35&1.63&1.82&1.91 \\ \hline
& 0.55&0.38&0.26&0.29&0.44&0.62&0.80 \\ 
(-0.50, -0.37) &$-164^{\circ}$&$-149^{\circ}$&$-115^{\circ}$&$-66.6^{\circ}$&$-41.3^{\circ}$&$-29.8^{\circ}$&$-23.4^{\circ}$ \\
&0.35&0.62&0.92&1.21&1.46&1.63&1.72 \\  \hline 
 &0.61&0.45&0.33&0.31&0.41&0.56&0.72 \\ 
(-0.38, -0.47) &$-164^{\circ}$&$-150^{\circ}$&$-125^{\circ}$&$-86.9^{\circ}$&$-56.7^{\circ}$&$-40.4^{\circ}$&$-31.2^{\circ}$ \\ 
&0.33&0.58&0.85&1.12&1.35&1.51&1.58 \\ \hline 
 &0.59&0.40&0.23&0.17&0.31&0.49&0.67\\ 
(-0.38, -0.27)&$-170^{\circ}$&$-162^{\circ}$&$-140^{\circ}$&$-79.1^{\circ}$&$-37.8^{\circ}$&$-24.1^{\circ}$&$-17.8^{\circ}$ \\ 
&0.31&0.55&0.81&1.07&1.28&1.44&1.51 \\ \hline
\end{tabular}
\end{center}
\caption{ Determination of $P/T$ (upper value), $\delta_{PT}$ (middle value) and $R$ (bottom value) 
using experimental results for $S_{\pi^+\pi^-}=(-0.50\pm 0.12)$ and $C_{\pi^+\pi^-}=(-0.37\pm 0.10)$ for given values of $\gamma$, $\gamma=(27^{\circ}\sim 87^{\circ})$ . } 
\end{table*}

 \begin{figure*}[t]
\begin{center}
\psfrag{ct}[c][c][1]{$C/T$}\psfrag{dct}[c][c][1]{$\delta_{CT}$}
\psfrag{rn}[c][c][.8]{\ $R_{00}$}\psfrag{rc}[c][c][.8]{\ $R_{+-}$}
\psfrag{gamma}[c][c][0.8]{$\gamma =$}
\includegraphics[width=5cm]{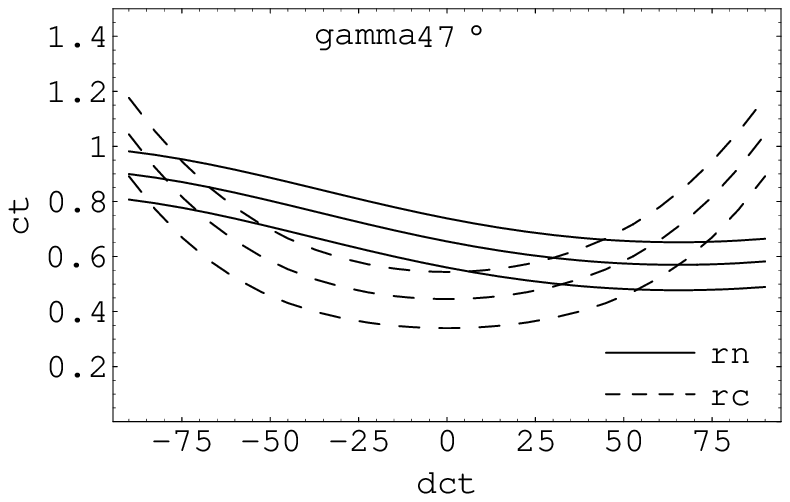}\hspace*{1cm}
\includegraphics[width=5cm]{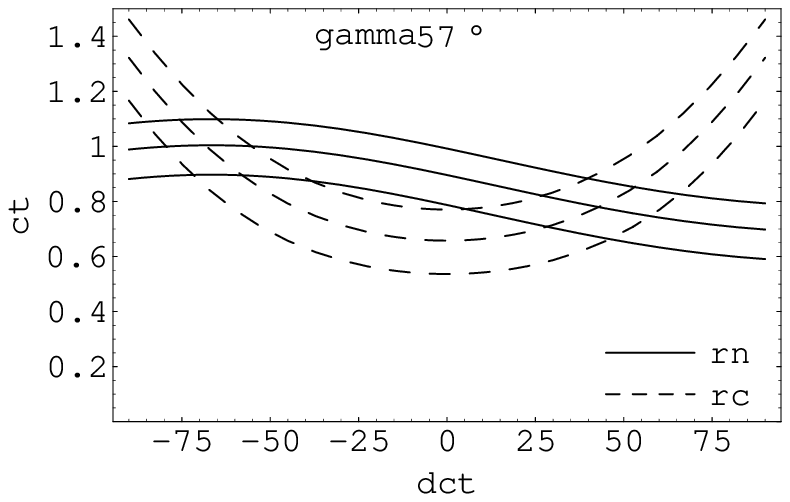} \vspace*{.5cm}\\
\includegraphics[width=5cm]{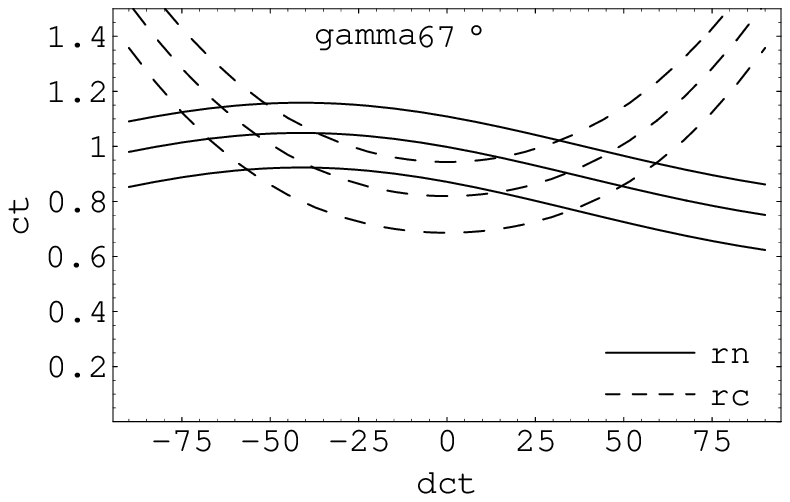}\hspace*{1cm}
\includegraphics[width=5cm]{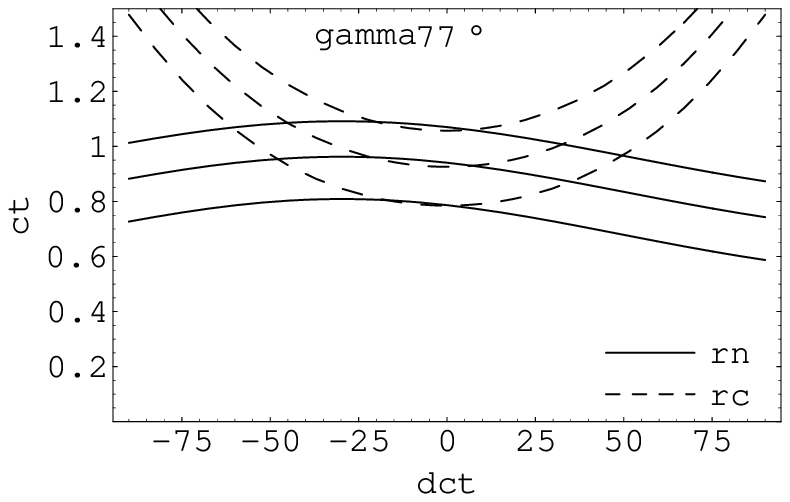}
\caption{ Allowed region for $\delta_{CT}$ ($x$-axis) versus
    $C/T$ ($y$-axis), from  bounds for $R_{00}$ and $R_{+-}$. Values for  $P/T$ and $\delta_{PT}$ are from the central value of  $(S_{\pi^+\pi^-}, C_{\pi^+\pi^-})=(-0.50, -0.37)$.
The three solid lines represent $R_{00}=0.64-0.14,\ 0.64, \ 0.64+0.14, $
the three dashed lines represent  $R_{+-}=2.27-0.32,\ 2.27, \ 2.27-0.32$.  
The overlap of solid and dashed bounds are the allowed region for $C/T$ 
and $\delta_{CT}$. }
\label{fig:r47}
\end{center}
\end{figure*}
\begin{figure*}[t]
\begin{center}
\psfrag{ct}[c][c][1]{$C/T$}\psfrag{dct}[c][c][1]{$\delta_{CT}$}
\includegraphics[width=5cm]{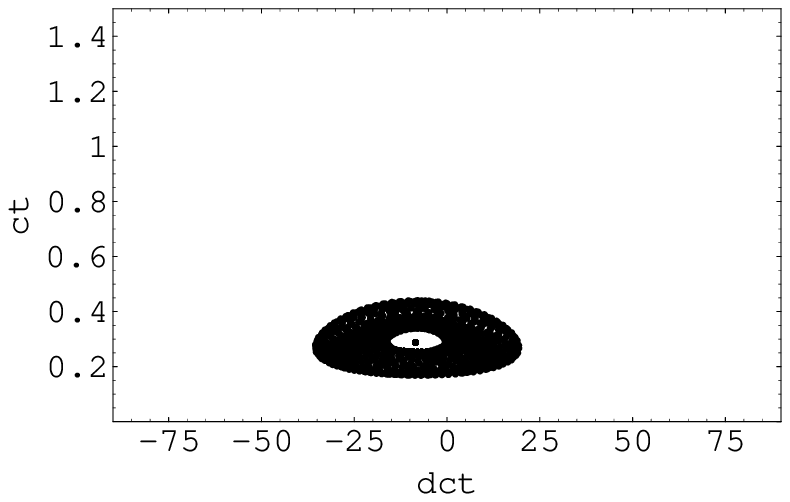}\hspace*{0.3cm}
\includegraphics[width=5cm]{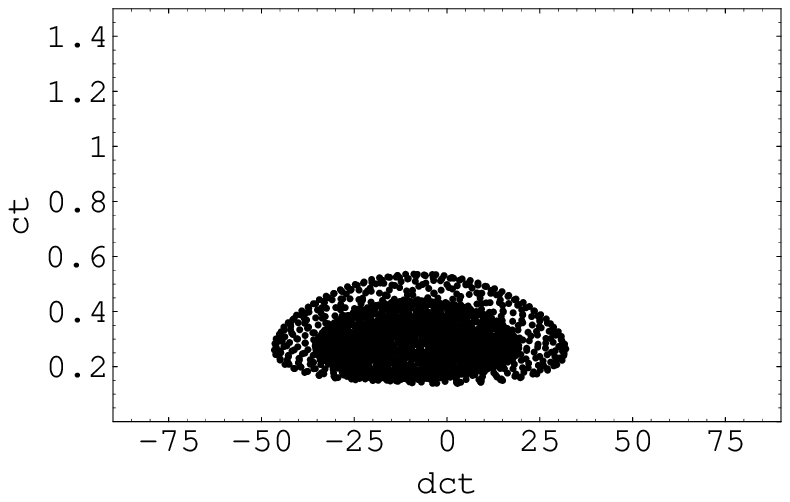}\hspace*{0.3cm}
\includegraphics[width=5cm]{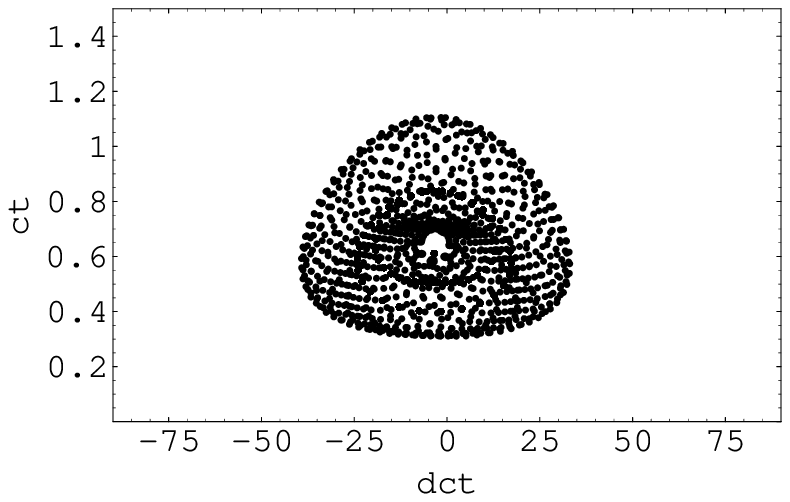}
\caption{Scattered plot of the QCD factorization estimate for $\delta_{CT}$ ($x$-axis) versus $C/T$ ($y$-axis)  including the end-point singularity effects. In the plot, we fix the $\rho$ parameters as $(\rho_H,  \rho_A)=(1, 1)$ (left) and $=(1, 2)$ (middle) and vary the phases in the range of  $-\pi < \phi_{A, H}<\pi$ (interval of 0.2 radian). The rest of the parameters are fixed (see text for details). The last figure (right) is obtained in the same manner with $(\rho_H,  \rho_A)=(1, 1)$ but with different parameter set, the so-called scenario 2 of QCD factorization (see text for details). }\label{fig:2}
\end{center}
\end{figure*}
\begin{figure*}[t]
\begin{center}
\psfrag{d0}[c][c][0.7]{$\delta_{20}=0^{\circ}$}
\psfrag{d20}[c][c][0.7]{$\delta_{20}=20^{\circ}$}
\psfrag{d40}[c][c][0.7]{$\delta_{20}=40^{\circ}$}
\psfrag{d60}[c][c][0.7]{$\delta_{20}=60^{\circ}$}
\psfrag{d80}[c][c][0.7]{$\delta_{20}=80^{\circ}$}
\psfrag{dm20}[c][c][0.7]{$\delta_{20}=-20^{\circ}$}
\psfrag{dm40}[c][c][0.7]{$\delta_{20}=-40^{\circ}$}
\psfrag{dm60}[c][c][0.7]{$\delta_{20}=-60^{\circ}$}
\psfrag{dm80}[c][c][0.7]{$\delta_{20}=-80^{\circ}$}\psfrag{ct}[c][c][1]{$C_{\rm eff}/T_{\rm eff}$}\psfrag{dct}[c][c][1]{$\delta_{CT_{\rm eff}}$}
\psfrag{rn}[c][c][.9]{$R_{00}$}\psfrag{rc}[c][c][.9]{$R_{+-}$}
\psfrag{gamma}[c][c][0.8]{$\gamma =$}
\includegraphics[width=5cm]{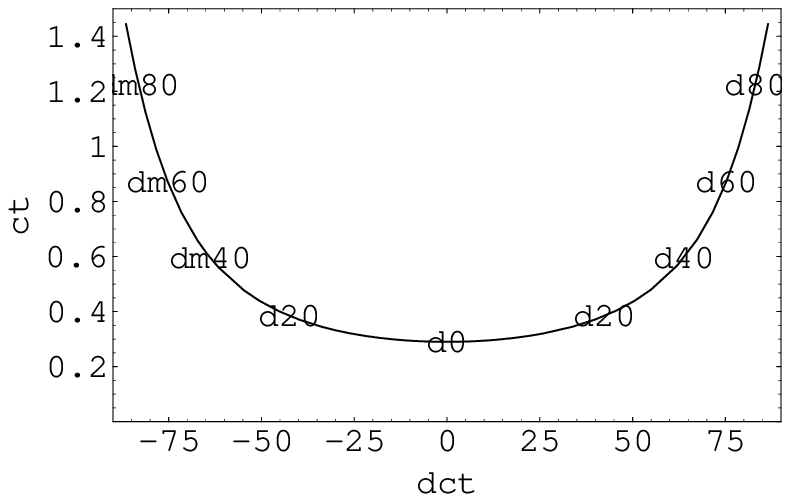}\hspace*{0.3cm}
\includegraphics[width=5cm]{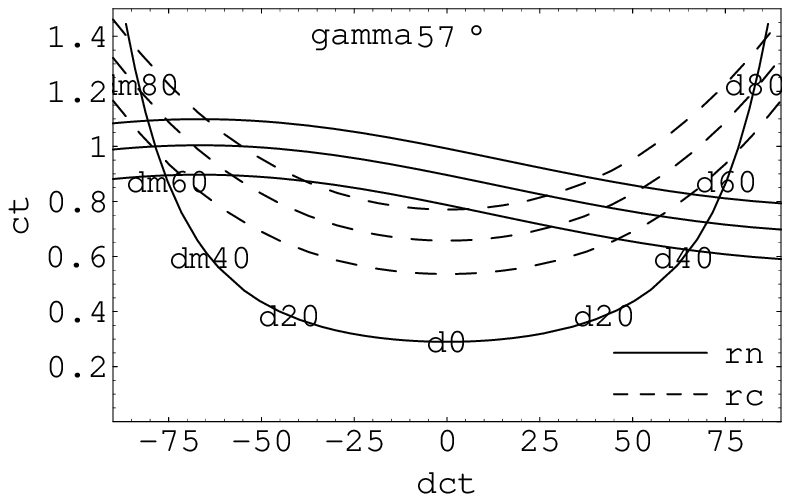}\hspace*{0.3cm}
\includegraphics[width=5cm]{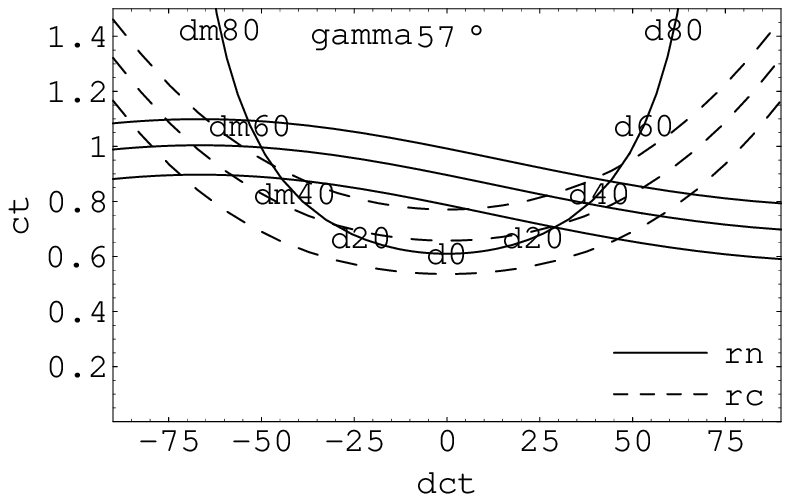}
\caption{The left figure is the plot of Eq. (\ref{eq:ct1}), the result including the FSI phase,  
in the plane of $\delta_{CT_{\rm eff}}$ ($x$-axis)
 versus  $C_{\rm eff}/T_{\rm eff}$ ($y$-axis) by varying $\delta_{20}$.
 versus  $C_{\rm eff}/T_{\rm eff}$ ($y$-axis) by varying $\delta_{20}$. 
 The number on the line indicates the value of $\delta_{20}$ at each point. 
The bare value $C_0/T_0=0.29$  obtained from the default parameter sets of the QCD factorization with $\rho_{H, A}=0$ is used.  In the middle figure, we put together the left figure and the experimental bounds from $R_{+-}$ and $R_{00}$ for the case of $\gamma=57^{\circ}$ (Fig.1 upper-right). The right figure is obtained in the same way as the middle one but using $C_0/T_0\simeq 0.61$, the result with  the parameter set called scenario 2 in QCD factorization. }\label{fig:3}
\end{center}
\end{figure*}

Figure \ref{fig:r47}  shows the allowed region for 
$C/T$ and $\delta_{CT}$ plotted with the central values of the CP
asymmetries measurements. Note that the overlap of the solid ($R_{00}$)
and the dashed ($R_{+-}$)  bounds shift towards the larger $C/T$ region 
as $\gamma$ becomes larger. $R_{+-}$ allows relatively small value of $C/T$ 
while $R_{00}$ leads to a more strict constraint, $C/T \gsim 0.5$.The
overlap region  is distributed in a large range of $\delta_{CT}$.   
Note also errors due to $(S_{\pi^+\pi^-}, C_{\pi^+\pi^-})$ are 
$\pm$ a few \% in $C/T$ and  $\pm 20^{\circ}$ in $\delta_{CT}$.
This analysis shows clearly that $C/T$ 
is large compared to the leading order QCD corrections($C/T=1/N_{c}$). 
In the next section
we present results of QCDF calculation for $C/T$.

\section{$C/T$ in QCD Factorization}
 Neglecting electroweak penguin contributions, $T$, $C$
and $P$ for $B\to \pi\pi$ decays in  QCD Factorization \cite{Beneke,BN}  
are given by:
\bea
Te^{i \delta_T}e^{i\gamma}&\propto&\lambda_u^*(a_1+b_1+\hat{a}^u_4) , \label{eq:Tqcd}\\ 
Ce^{i \delta_C}e^{i\gamma}&\propto&\lambda_u^*(a_2-b_1-\hat{a}^u_4) , \label{eq:Cqcd}\\ 
Pe^{i \delta_P}&\propto&\lambda_c^*\hat{a}^c_4 \label{eq:Pqcd}
\eea
\be
\hat{a}_4^p=a_4^p+r_\chi a_6^p+2b_4+b_3
\ee
$r_\chi=2m_{\pi}^2/(2m_b m_q)\simeq 1.24 $ and 
 $m_q\equiv (m_u+m_d)/2$ . 
\bea
\lambda_u=V_{ub}V_{ud}^*\simeq A\lambda^3(\rho -i\eta ) \label{eq:lambdau}\\
 \lambda_c=V_{cb}V_{cd}^* \simeq -A\lambda^3. \label{eq:lambdac}
\eea
where $a_{i}$ contain the factorisable term, the vertex correction, 
hard-scattering correction and penguin correction and $b_{i}$ are
annihilation contributions.
Using the following parameters,
\bea
&\mu=4.2\,\rm GeV,\quad  m_q(2\rm GeV)=0.0037\,\rm GeV ,\quad
 \lambda_B=0.35\,\rm GeV , \nonumber \\ 
&|\lambda_u/\lambda_c|=|\rho +i\eta |= 0.09, \quad \alpha_2^{\pi}=0.1 
\eea
we find 
\bea
 a_1&=&1.02 e^{i 0.8^{\circ}}-0.014 \rho_H e^{i \phi_H} \label{eq:a1} \\ 
 a_2&=&0.21e^{-i 23^{\circ}}+0.081  \rho_H e^{i \phi_H} \label{eq:a2} \\
 a_4^u+r_\chi a_6^u&=&
- 0.097e^{i 21^{\circ}}+0.0010  \rho_H e^{i \phi_H}\label{eq:a4u} \\
a_4^c+r_\chi a_6^c&=&
- 0.10e^{i 7^{\circ}}+0.0010  \rho_H e^{i \phi_H}  \label{eq:a4c}
\eea
and 
\bea
 && b_1=0.027+0.063\rho_A e^{i\phi_A} +0.0085(\rho_A e^{i \phi_A})^2  \label{eq:b1} \\
 && b_3=-0.0067-0.021\rho_A e^{i\phi_A} -0.015(\rho_A e^{i \phi_A})^2  \label{eq:b2} \\
 &&b_4=-0.0019-0.0046\rho_A e^{i\phi_A} -0.0006(\rho_A e^{i \phi_A})^2 \label{eq:b3} 
\eea
We have made a complete analysis of $C/T$ covering all 
the parameter space of $\rho$'s and $\phi$'s. In the limit of 
$\rho_{H, A}=0$ . We find
\be
\frac{C_0}{T_0}e^{i\delta_{CT_0}}= 0.29 e^{-i8.5{\circ}} \label{eq:ctrho0}
\ee
$\rho_{H, A}$ and $ \phi_{A, H}$ represent the divergent end-point
singularity terms 
for the hard scattering function and the annihilation diagrams which
introduce large theoretical uncertainties in the QCD calculation.  
 Fig.\ref{fig:2} 
shows $\delta_{CT}$ versus $C/T$
for $-\pi < \phi_{A, H}<\pi$ with fixed $\rho_{H}$ and $\rho_{A}$ which
would   vary, say, in the ranges of  $|\rho_{A,  H}|<1\sim 2$. We see that
quite a large range of $C/T$ and $\delta_{CT}$ are allowed from 
QCD factorization, $C/T$  up to 0.45 (0.55) for $\rho_A=1 (2)$. However 
the range of allowed values for
$C/T$ is still below the fitted values  shown in  Figure \ref{fig:r47}
for $\gamma > 47^{\circ}$ . We thus obtain a constraint   
$\gamma \le 44^{\circ} (52^{\circ})$  and $\gamma \le 46^{\circ}$ 
($56^{\circ}$),  for  $\rho_A=1$ and $\rho_A=2$ respectively.
In \cite{BN} the problem of the small $a_2$  has already been 
 recognized and a possible solution was proposed : by choosing the
the largest value of the Gegenbauer moment of $\pi$ distribution amplitude
(scenario S2). More recently, this model   
 has been reanalyzed  with NLO corrections to the 
 hard spectator-scattering diagram \cite{beneke2} and an enhancement
by a factor of 2 for the hard-scattering correction is obtained:
\be
a_2\simeq 0.48 e^{-i10^{\circ}}+0.18 \rho_H e^{i \phi_H} \label{eq:s2}
\ee
\be
C_0/T_0 e^{i\delta_{CT_0}}\simeq 0.61 e^{-i3^{\circ}} , \rho_{H, A}=0.
\ee
As shown in the scattered plot of Fig. 2 with $\rho_A=1$ for scenario 
S2(right most),
$C/T\simeq 1.1$ can be achieved in this scenario if 
$\delta_{CT}$ is very small. It appears that QCD factorisation can 
solve the large $C/T$ puzzle. 

\section{Does FSI Phase Make $C/T$ large}
Recent studies show that  FSI phase in $B\to \pi\pi$
can be large\cite{Suzuki}
and could effectively enhances $C/T$\cite{Cheng,Fajfer}, since 
the $B\to \pi^{0}\pi^{0}$ decay can
 be induced by the charge exchange scattering process $\pi^{+}\pi^{-}\to
\pi^{0}\pi^{0}$ which  can  generate  $C$ 
from $T$. We examine here whether the FSI effect  can enhance
sufficiently the value 
of $C/T$ of the QCD factorization 
without adjusting the incalculable parameters $\rho_{H,A}$ and $\phi_{H,A}$.
We now add the FSI phases to  the QCDF amplitude and evaluate $C/T$
and constrain the values of $\delta_{20}=\delta_{2}- \delta_{0}$  
and $\delta_{CT}$ using the 
allowed region for $C/T$ in Fig.1 with $\gamma = 57^{\circ}$. In \cite{WW}, a
similar analysis is performed and a large
$\delta_{20}$ is found by a fit to the the central values of the
experimental data. We have
\bea
 &&T_{\rm eff} e^{i\delta_{T_{\rm eff}}} = [(2T_0-C_0)e^{i\delta_0}+ (T_0+C_0)e^{i\delta_2}]/3 
 \nn\\
 &&C_{\rm eff} e^{i\delta_{C_{\rm eff}}} = [-(2T_0-C_0)e^{i\delta_0}+ 2(T_0+C_0)e^{i\delta_2}]/3 \nn \\
 &&P_{\rm eff} e^{i\delta_{P_{\rm eff}}}=P_{0}e^{i(\delta_D+\delta_{0P})}. \label{eq:29}
\eea
where  $C_0, T_0, P_0$ are $C$, $T$ and $P$ for $\rho_{H, A}=0$
\be
\left(\frac{C_{\rm eff}}{T_{\rm eff}}\right)e^{i\delta_{CT_{\rm eff}}}
=\frac{(-2+2e^{i\delta_{20}})+(1+2e^{i\delta_{20}})C_0/T_0}{(2+e^{i\delta_{20}})+(-1+e^{i\delta_{20}})C_0/T_0}\ \ \ \label{eq:ct1}
\ee
$(C/T)_{\rm eff}$ indeed becomes larger as the
FSI phase increases. The bare ratio
 $C_0/T_0=0.29$ can be enhanced to $(C_{\rm eff}/T_{\rm eff})\simeq 0.4$
  for $\delta_{20}\simeq \pm 21^{\circ}$, where 
$\delta_{CT_{\rm eff}}\simeq \pm 44^{\circ}$.  

 For $\gamma=57^{\circ}$, $C_0/T_0=0.29$, the allowed
 region from $R_{+-}$ and 
$R_{00}$ overlap at $\delta_{20}\simeq -65^{\circ}$, at which 
$(C/T)_{\rm eff}\simeq 0.96$ (middle figure). 
 For $\gamma=57^{\circ}$, 
$C_0/T_0=0.61$(scenario S2), the central value of $(R_{+-}, R_{00})$ 
are reproduced by 
$\delta_{20}\simeq 40^{\circ}$ where $C_{\rm eff}/T_{\rm eff}\simeq 0.8$ and $\delta_{CT_{\rm eff}}\simeq 40^{\circ}$. 
 To  constrain $\gamma$, we need the FSI phase and input from QCDF. Then
 assuming $\delta_{20}\lsim 30^{\circ}$, we obtain
$\gamma \lsim 48^{\circ} (55^{\circ})$ using the QCDF default  values. 
With the parameters for  scenario 2 shown above,   we find 
that  $\delta_{20}\lsim
 30^{\circ}$ leads to  $\gamma = (59\pm3)^{\circ} (>56^{\circ})$. Note
 that there may be a FSI contribution not only to
the phase but to $C/T$ itself\cite{Fajfer}, so the bounds obtained above may
receive  corrections from QCD or FSI effects.
\section{Conclusion}
Using the latest $B\to\pi\pi$ data, we find that in $B\to \pi\pi$ 
decays $C/T$ is large  over a  large range for
$\delta_{CT}$: $C/T \gsim 0.5$ for $\gamma > 47^{\circ}$. 
We then show that QCDF with  large $\rho_{H, A}$ leads to large
values of $C/T$, especially when $\delta_{CT}$ is small. In  a scenario 
of QCDF with  $\rho_{H, A}=0$ but with the FSI strong phase included,   
$C/T$ and $\delta_{CT}$ are enhanced when the FSI phase $\delta_{20}$ 
increases. A precise measurement of the $B^{0}\to \pi^{0}\pi^{0}$
direct CP asymmetry $C_{00}$ could distinguish these two models. In fact
\be
\frac{C_{00}}{C_{+-}}=\frac{C}{T}\frac{\sin (\delta_{CT}-\delta_{PT})}{\sin \delta_{PT}} \frac{1}{R_{00}}. 
\ee 
 A small $\delta_{CT}(\simeq 0)$ leads to a   ratio of order 
unity with negative sign,  $C_{00}/C_{+-}\simeq -(C/T)/R_{00}$, implying
$C_{00}=0.57$ for $\delta_{CT}=0$ , close to the
 higher end of the current experimental value 
 $C_{00}= 0.28^{+0.40}_{-0.39}$ \cite{HFAG} while a large  
$\delta_{CT}(\simeq \pm \pi/2)$ would imply  a strong 
dependence on $\delta_{PT}$ for the ratio
$C_{00}/C_{+-}\simeq \pm (C/T)/(R_{00}\tan\delta_{PT})$.

\section{Acknowledgments}
I would like to thank Giulia Ricciardi and The Organizers for the Invitation 
to The Workshop and for the warm hospitality extended to me at Capri.


\end{document}